\documentclass[doublecol]{epl2}
\usepackage{graphicx}
\usepackage{amsmath}
\usepackage{amssymb}

\DeclareMathOperator{\Tr}{Tr}

\renewcommand{\Re}{\mathop\mathrm{Re}\nolimits}

\def\mathbold{\bf}
\def\be{\begin{equation}}
\def\ee{\end{equation}}
\def\bea{\begin{eqnarray}}
\def\eea{\end{eqnarray}}
\def\br{{\mathbold r}}

\title{Nonequilibrium mesoscopic superconductors in a fluctuational regime}
\shorttitle{Nonequilibrium Ginzburg-Landau-type theory}

\author{N. Chtchelkatchev \inst{1,2} \and V. Vinokur \inst{2} }
\shortauthor{N. Chtchelkatchev \etal}
\institute{
\inst{1} Department of Theoretical Physics, Moscow Institute of Physics and Technology, 141700 Moscow, Russia

\inst{2} Argonne National Laboratory, Argonne, IL 60439, USA
 }


\abstract{
We develop a non-equilibrium Ginzburg-Landau type theory of the far from the equilibrium dynamics of superconductors in a fluctuational regime and apply our approach to quantitative description of a superconductor island in a stationary nonequilibrium state.  We derive the effective temperature of the nonequilibrium state and find fluctuational contributions to the magnetic susceptibility showing that it becomes a singular function of $\sqrt{V-V_{\rm c}}$, where $V$ is the external drive and $V_{\rm c}$ is its ``critical'' value at which the nonequilibrium phase transition takes place.
}

\begin{document}

\maketitle

Time-dependent Ginzburg-Landau equation (TDGL) successfully describes
\textit{weakly} non-equilibrium dynamics of the systems experiencing a second order
phase transition, including the ferromagnetic, superconducting, and the superfluid
transitions, to name a
few~\cite{Gorkov,Eliashberg68,NonequilibriumSuperconductivity,Kopnin_book,Schmid,Larkin_Ovchinnikov,Kramer,Stoof,Levchenko-Kamenev,Nazarov},
in the vicinity of the critical point.
A marked progress has been recently achieved~\cite{Levchenko-Kamenev} in formulation
Keldysh technique-based approach aimed at extension of TDGL onto strongly
non-equilibrium situation,~see, e.g. the review Ref.~\cite{Kamenev_review}.  Yet
constructing a theory of far from the equilibrium dynamics for the GL systems
remains a major challenge of the nonequilibrium statistical physics.

A nonequilibrium extension of the Ginsburg-Landau theory (NGL) requires, in
principle, the details of the underlying microscopic physics.  The latter enter the
theory through the quasiparticle density of states and relaxation rates
that appear as parameters in corresponding kinetic equations (KE) for quasiparticle
distribution functions which, in their turn, depend on the order parameter $\Delta$
\cite{NonequilibriumSuperconductivity,Kopnin_book}.
However, as we show below, one can construct a phenomenological non-equilibrium
theory in a critical region without invoking the details of the microscopic behavior
of a low-symmetry phase, making use of the symmetry considerations in a spirit of
Ref.~\cite{Landau_theory}.

In this Letter we develop a theory of the \textit{far from-equilibrium} fluctuation
effects in superconductors
generalizing a standard description of superconducting
fluctuations~\cite{Larkin-Varlamov_book,Levchenko-Kamenev} to the far from the
equilibrium state and derive a closed GL-like equations set, describing
quantitatively far from the equilibrium dynamics of fluctuations.  In particular, we
calculate the nonequilibrium fluctuation contribution to magnetic susceptibility and
find an analytical expression for the effective temperature, $T_{\rm
eff}=T\cosh^2(V/4T)$, depending on the bias, $V$. As an illustration of the proposed
general technique, we discuss the dynamics of a superconducting granule connected to
reservoir via disordered normal wires.

In the framework of a phenomenological theory of a second order phase transition,
the properties of the low-temperature phase near the transition are determined by
the free energy functional written as an expansion in the order parameter, see, e.g,
Refs.~\cite{Landau_theory,Ginsburg-Landau}.
In an equilibrium, the density matrix space $\mathbb{M}$ is parameterized by the
temperature and the transition occurs at $T=T_{\rm c}$; the corresponding
dimensionless parameter of the Landau expansion is $|\tau_{\rm {\scriptscriptstyle
GL}}^{(eq)} T_{\rm c}|^{-1}\sim|T-T_{\rm c}|/T_{\rm c}\ll 1$ (we will use the units
where $\hbar=k_B=e=1$ throughout the paper), where $\tau_{\rm {\scriptscriptstyle
GL}}^{(eq)}$ is the Ginzburg-Landau relaxation time in an equilibrium.  In a general
nonequilibrium case the transition extends over some \textit{surface}  in
$\mathbb{M}$ (equilibrium density matrices form a zero measure subspace of
$\mathbb{M}$).

We consider a system where the excitations and their kinetics are well defined. Then
a nonequilibrium GL theory has the form similar to that of the equilibrium one,
provided the distribution functions of the excitations are stationary. Following the
general recipe~\cite{Kamenev_review} for treating an out of the equilibrium system,
one is to use the Keldysh partition function, instead of the conventional partition
function for the equilibrium case, and, accordingly, the Keldysh action for the
order parameter replaces the GL free energy in equilibrium. The parameters of the
nonequilibrium GL-expansion are functionals of the excitation distribution
functions. The closeness of the system to the (non-equilibrium) phase transition
surface is determined by the dimensionless parameter $|\tau_{\rm {\scriptscriptstyle
GL}}T_{\rm eff}|^{-1}\ll 1$, where $T_{\rm eff}$ is the nonequilibrium energy scale
replacing $T$ in an out of the equilibrium state and $\tau_{\rm {\scriptscriptstyle
GL}}$ is a general, nonequilibrium, GL relaxation time replacing $\tau_{\rm
{\scriptscriptstyle GL}}^{(eq)}$. {This inequality is the necessary condition for applicability of our approach.}
Note that the existence of $T_{\rm eff}$ does not imply the local equilibrium form of the excitation distributions.

The coefficients of the nonequilibrium GL functional behave differently as compared
to those of the equilibrium. In particular, the coefficient at the forth order term
in the order parameter, $\Delta$, can even change the sign at large driving forces.
This signals the onset of an instability  of  the NGL-equations solution and means
that the NGL functional should be expanded to higher orders with respect to
$\Delta$.  In the context of superconductivity, the equilibrium GL-expansion has the
usual form~\cite{Ginsburg-Landau}:
\begin{gather}\label{eq:Omega_superconductivity}
\Omega[\Delta]=\nu a \int_0^\beta d\tau d\br\left\{ \Delta^* L^{-1}\Delta+\frac b {2T}|\Delta|^4\right\},
\end{gather}
where $L^{-1}=i\partial_\tau+(\tau_{\rm {\scriptscriptstyle GL}}^{(\rm
eq)})^{-1}-D\partial_{\mathbf{r}}^2$, $(\tau_{\rm {\scriptscriptstyle GL}}^{(\rm eq)})^{-1}=(T-T_{\rm c})\,\alpha$, $Z_\Delta=\int D\Delta D\Delta^*
\exp\{-\Omega[\Delta,\Delta^*]\}$ is the partition function, and $\nu$ is the density of states (DoS) at the Fermi shell. In a  disordered metal model $a=\pi/8 T$, $\alpha=8/\pi$, $D$ is the diffusion coefficient and $b=7\zeta(3)/\pi^3$ \cite{Gorkov}.

The order parameter in the Keldysh space has two components, corresponding to the lower and upper brunches of the Keldysh time-contour. {To simplify the structure of the action, it is convenient to use the rotated basis and introduce ``classical'' ($\Delta_1$) and ``quantum'' ($\Delta_2$) components of the order parameter [half-sum and half-difference of the order parameter values at the lower and upper brunches of the Keldysh time-contour]\cite{Levchenko-Kamenev,Keldysh}. Thus $\overrightarrow{\Delta}=(\Delta_1, \Delta_2)^\tau$.} The Keldysh analog of the partition function $Z_\Delta$ is
\begin{gather}
\mathcal Z_\Delta=\int D\overrightarrow{\Delta} D\overrightarrow{\Delta}^*\exp\{iS_\Delta[\overrightarrow{\Delta},\overrightarrow{\Delta}^*]\}.
\end{gather}
The average order parameter should be identified with $\langle\Delta_1\rangle$, while $\langle\Delta_2\rangle=0$, where averaging is based on $\mathcal Z_\Delta$:
\begin{gather}
{\langle\ldots\rangle=\int D\overrightarrow{\Delta} D\overrightarrow{\Delta}^*[\ldots]\exp\{iS_\Delta[\overrightarrow{\Delta},\overrightarrow{\Delta}^*]\}.}
\end{gather}
{Here we used the fact that in the absence of the quantum components of external source-fields, $\mathcal Z_\Delta\equiv 1$, this is a manifestation of the causality principle. The saddle point equation}
\begin{gather}\label{eq:1223}
{\frac{\delta S_\Delta[\overrightarrow{\Delta},\overrightarrow{\Delta}^*]}{\delta\Delta^*_2}=0,}
\end{gather}
{at the manifold $\Delta_2^*=\Delta_2=0$ produces the nonequlibrium generalization of the GL equations.}

\begin{figure}[t]
  \includegraphics[width=85mm]{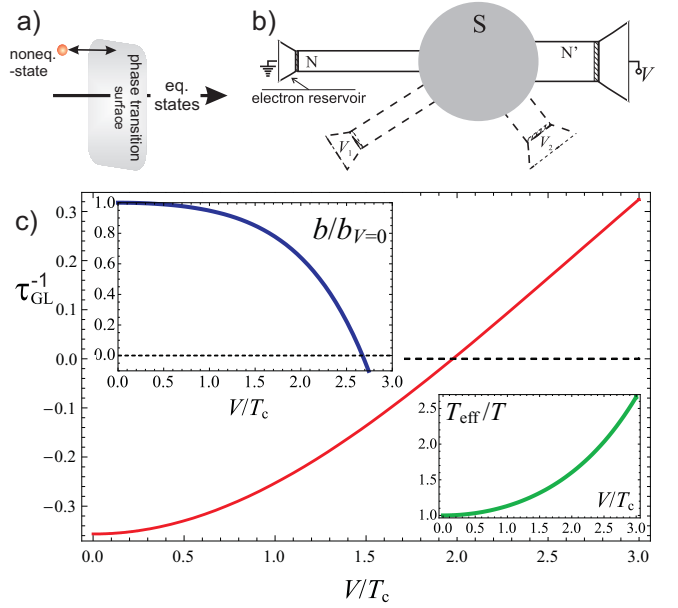}\\
  \caption{
  (color online) 
 a) Density matrix space. b) An exemplary system: superconducting granule weakly
connected to the reservoirs.  c)  The GL nonequilibrium relaxation time as a
function of the applied voltage $V$ at the resevoir temperature $T=0.7T_{\rm c}$.
[The units of $\tau_{\rm {\scriptscriptstyle GL}}^{-1}$ are chosen to match
$\ln(T/T_{\rm c})$ in an equilibrium].  The lower inset illustrates that $T_{\rm
eff}$ differs from $V$. The upper inset shows $b(V)$ and illustrates the difference
between the nonequilibrium and equilibrium behaviors; in the latter case $b$ is the
temperature independent constant. }\label{fig3}
\end{figure}
{Near the phase transition we can expand $iS_\Delta[\overrightarrow{\Delta},\overrightarrow{\Delta}^*]$ over $\Delta$. This expansion over quantum components of the order parameter  should be performed up to the first order, see Eq.\eqref{eq:1223}.} Carrying out the microscopic calculation based on the Keldysh functional representation  of the BCS theory \cite{BCS} in a form of the nonlinear $\sigma$-model we find that the shape of $S_\Delta[\overrightarrow{\Delta},\overrightarrow{\Delta}^*]$ above and below [the restrictions are discussed] the phase transition can be written in a form similar to that of
\eqref{eq:Omega_superconductivity}:
\begin{gather}\label{eq:cZ_superconductivity}
S_\Delta=  {\nu a}\Tr\left\{\overrightarrow{\Delta}^\dag\left[\hat
L^{-1}-\frac{b}{T_{\rm eff}} |\Delta_1|^2\hat\sigma_x\right]\overrightarrow{\Delta} \right\},
\end{gather}
{where  $\Tr$ means the trace with respect to times and the integration over
coordinates, while $\hat\sigma_x$ is the Pauli matrix in the Keldysh space. We show below that Eq.\eqref{eq:cZ_superconductivity} can be derived as well from the symmetry considerations similar to those of Ref.\cite{Landau_theory}.}

The microscopic Keldysh nonlinear $\sigma$-model calculation expresses GL coefficients $a$ and $b$ through the gauge invariant  electron and hole distribution functions~\cite{Kopnin_book}, $\tilde F_{\rm e}$ and $\tilde F_{\rm h}$ [or
$\tilde F_{\rm L}=(\tilde F_{\rm e}+\tilde F_{\rm h})/2$ and $\tilde F_{\rm T}=(\tilde F_{\rm e}-\tilde F_{\rm h})/2$], where in an equilibrium, $\tilde F_{\rm {e(h)}}=\tanh(\epsilon/2T)$:
\begin{gather}\label{eq:a}
{a=\frac \pi4 \lim_{\epsilon\to0}\partial_\epsilon \tilde F_{\rm L}(\epsilon)\equiv\frac{\pi}{8T_{\rm eff}},}
\\\label{eq:b0}
    {b=4T_{\rm eff}^2\int_0^\infty t \tilde Y(t)dt,}
\end{gather}
where $2\pi\tilde Y(t)=\int_0^\infty d\epsilon \exp[-iEt]\tilde F_{\rm L(E)}/(E+i0)$. {One can see straightforwardly that in the equilibrium limit Eqs.\eqref{eq:a}-\eqref{eq:b0} reproduce conventional values $a=\pi/8T$ and $b=7\zeta(3)/\pi^3$ respectively. Formulas \eqref{eq:a} and \eqref{eq:b0} for the coefficients of the far from equilibrium GL functional constitute the main result of our work.}

{A structure of $\hat L$ in a Keldysh space is determined by the causality principle\cite{Kamenev_review} and by the comparison with the standard form of GL expression given by Eq.\eqref{eq:Omega_superconductivity}}
\begin{gather}
\hat L^{-1}=\begin{pmatrix}
              0 & L_{\rm A}^{-1} \\
              L_{\rm R}^{-1} & [L^{(-1)}]_{\rm K} \\
            \end{pmatrix},
\end{gather}
{where $R(A)$ and $K$-subscripts denote the retarded (advanced) and Keldysh propagators respectively.} The form of the $R(A)$-component can be found from the correspondence between the imaginary and real time representations that can be transformed into each other by the Euclidean rotation:
\begin{gather}
{L_{\rm {R(A)}}^{-1}=\pm(\partial_{\rm t}-2i\varphi)-\tau_{\rm {\scriptscriptstyle GL}}^{-1} -D(\nabla_{\mathbf{r}}-2i\mathbf{A})^2,}
\end{gather}
where $(\varphi,\mathbf{A})$ is the electromagnetic potential  \cite{footnote:imbalance}].

The Keldysh component of the fluctuation propagator is expressed via the additional temperature scale $T^*$:
\begin{gather}\label{eq:T_second}
{[L^{(-1)}]_{\rm K}\equiv4iT^*\,,}
\end{gather}
where $T^*=[1-\tilde F_{\rm e}(0)\tilde F_{\rm h}(0)]T_{\rm eff}$ is determined from the microscopic calculation and
$\tau^{-1}_{\rm {\scriptscriptstyle GL}}$ is determined by (provided the time reversal symmetry is not broken):
\begin{gather}\label{eq:tau_GL_def_final}
\tau_{\rm {\scriptscriptstyle GL}}^{-1}=T_{\rm eff}\int_\epsilon
K^{(L)}(\epsilon)\left[\tilde F_{\rm L}(\epsilon)-\tilde F_{\rm
L}^{(0)}(\epsilon)\right],
\end{gather}
where $K^{(L)}(\epsilon)=\alpha/2\epsilon$. Here $\tilde F_{\rm L}^{(0)}(\epsilon)$ is the distribution function at the phase transition surface.

The current can be found analogously to Ref.~\cite{Levchenko-Kamenev} by adding to the action the quantum
source in a form of the vector potential, $\mathbf{A}_2$, in addition to the
(classical) external fields potentials $(\varphi,\mathbf{A})$ discussed above, and
then varying the term proportional to $\mathbf{A}_2$ in $S_\Delta$. Thus, yields the
supercurrent as,
\begin{gather}
{\mathbf{j}_s=\pi e\nu D|\Delta_1|^2\partial_{\mathbf{r}}\arg(\Delta_1)/2T_{\rm eff}.}
\end{gather}

The conditions that electromagnetic fields and quasiparticle distribution functions are
stationary and weakly depend on coordinates on the scale of the Cooper pair size,
$\xi=\sqrt{D\tau_{\rm {\scriptscriptstyle GL}}}$, ensure the applicability of the
NGL-functional. We verify that our highly NGL equations reduce to the
standard \textit{weakly}-nonequilibrium form in the case where the deviations from the
equilibrium are small.  Indeed, if $\tilde F$ differs slightly from
$\tanh(\epsilon/2T)$, one has to use the equilibrium values for $a$ and $b$ and keep
nonequilibrium $\tau_{\rm {\scriptscriptstyle GL}}$ [since the nonequilibrium
corrections are of the next order if the deviation from the equilibrium is small]. This
means that our theory is restricted to the neighborhood of the intersection of the phase
transition surface with the equilibrium density matrix subspace, see Fig.\ref{fig3}.
Rewriting Eq.\eqref{eq:tau_GL_def_final} as $1/(\tau_{\rm {\scriptscriptstyle
GL}}T_{\rm eff})=\alpha (T-T_c)+\alpha\int_\epsilon [\tilde
F_L(\epsilon)-\tanh(\epsilon/2T)]/2\epsilon$  we recover the weakly-nonequilibrium
version of the GL theory \cite{Larkin_Ovchinnikov,Eliashberg68}.


Most of the quantities related to superconducting fluctuations are the singular functions of $\tau_{\rm {\scriptscriptstyle GL}}^{(\rm eq)}$ in the equilibrium~\cite{Larkin-Varlamov_book}. When we move out of the equilibrium, the fluctuation corrections are parameterized by the nonequilibrium $\tau_{\rm {\scriptscriptstyle GL}}$ and depend on $T_{\rm eff}$. Recently zero-dimensional superconducting fluctuations and fluctuating diamagnetism in the lead nanoparticles were experimentally investigated, see, e.g., Ref.\cite{Bernardi}.{ Motivated by the experiments we find, as an example,  the nonequilibrium, $V>V_{\rm c}$, fluctuation contribution to the magnetic moment of a small (of the size $L\ll\xi$) superconductor:}
\begin{gather}\label{eq:xi}
M=-2T_{\rm eff}\tau_{\rm {\scriptscriptstyle GL}} H D L^2\eta,
\end{gather}
where $H$ is the magnetic field and $\eta=1/10$ for a spherical island. Nonequilibrium fluctuation corrections to other quantities, e.g, related to the diffusion propagator contributions to $\mathcal Z$  [and the Langevin noise corrections with the correlator proportional $T^*$], will be presented in \cite{PRBvariant}.

Now we sketch a general procedure for calculating fluctuation-related quantities.
In an out of the equilibrium state, one uses the Keldysh real time partition function instead of the conventional thermodynamic partition function~\cite{Levchenko-Kamenev}:
\begin{gather}
{\mathcal Z=\int D[\mathcal A,\Psi,\bar\Psi {\Delta} ,{\Delta}^*] \exp\{iS\},}
\end{gather}
where the Grassman fields, $\Psi$, $\bar \Psi$, describe the fermion (superconductivity-related) degrees of freedom (on Keldysh contour), $\mathcal A=(\varphi,\mathbf{A})$ and $S[\mathcal A,\Psi,\bar\Psi {\Delta} ,{\Delta}^*]$ is the microscopic Keldysh action of the system. The Ginsburg-Landau expansion of the effective thermodynamical potential is an example of the so-called low energy field theory, i.e. a theory, where the order parameter fields change negligibly on the microscopic scales, e.g., the lattice constant. The low energy theory in Keldysh formalism appears after integrating out the high energy part of the fields. The resulting low energy effective action consists of the three parts. The first one, $S_\Delta$, describes quantum dynamics of the $\Delta$-field, the second part generates the kinetic equations for the excitations. It looks schematically like $\Tr[z\circ\,(\rm KE)]$, where the dynamic variable $z$ is closely related to the anti-Keldysh  component of the $Q$-matrix in the nonlinear $\sigma$-model formalism. Integrating over $z$ we obtain the functional $\delta$-function ensuring that the distribution function obeys the kinetic equations. The third part $S_A$ describes electromagnetic fields.   The variation of the effective action over $S_A$ produces Maxwell equations.

The fluctuations of the $\Delta$-field enter the collision integrals of the kinetic equations and the collisionless terms [the fluctuation renormalizations of the KE-coefficients], while $\langle\Delta_1 \rangle$ enter the (nonlinear) kinetic equations as external fields~\cite{Kopnin_book,PRBvariant}. The $\Delta$-fluctuations in KE contribute to the fluctuation corrections to the kinetic coefficients \cite{PRBvariant}. The $\langle\Delta_1\rangle$-terms in KE are important while $\tilde F_\epsilon$ differs essentially from $\tanh(\epsilon/2T)$ only at small energies, $\epsilon\sim\langle\Delta_1 \rangle$. But if quasiparticles are excited in the wide energy range above the gap, $\langle\Delta_1 \rangle\lesssim|\epsilon|\lesssim T_{\rm eff}$, then $\langle\Delta_1 \rangle$-terms in KE induce small, $\sim o(\langle\Delta_1\rangle/T_{\rm eff})\ll1$, the perturbation of $\tilde F_\epsilon$ and the subleading [$\sim\langle\Delta_1\rangle/T_{\rm eff}\ll1$] terms in $S_\Delta$  compared to the terms given in Eq.\eqref{eq:cZ_superconductivity}.


The phenomenological Landau theory predicts $(\tau_{\rm {\scriptscriptstyle GL}}^{\rm eq})^{-1}$ to depend linearly on $|T-T_{\rm c}|$ and $|\tau_{\rm {\scriptscriptstyle GL}}^{\rm eq}| T_{\rm c}\gg 1 $. In the nonequlibrium state the role of $(T-T_{\rm c})$ is taken by some functional of the electron and hole distribution functions, which characterizes  the effective ``distance" from the phase transition. We expect that $\tau_{\rm {\scriptscriptstyle GL}}^{-1}$ is a linear functional of {$\delta\tilde F_{\rm {e(h)}}=\tilde F_{\rm {L(T)}}-\tilde F_{\rm {L(T)}}^{(0)}$, otherwise the contribution proportional to $(T-T_{\rm c})^3$ to $\tau_{\rm {\scriptscriptstyle GL}}^{-1}$ would have appeared in an equilibrium. We write thus
\begin{gather}
{\tau_{\rm {\scriptscriptstyle GL}}^{-1}=T_{\rm eff}\int_\epsilon \{
K^{(L)}_\epsilon\delta\tilde F_{\rm L}(\epsilon)+
K^{(T)}_\epsilon\delta\tilde F_{\rm T}(\epsilon)\},}
\end{gather}
where the kernels $K$ are some functions of the energy. In an equilibrium $\tilde F_{\rm T}=0$ and $\tilde F_{\rm L}=\tanh(\epsilon/2T)$.} Then we reproduce the equilibrium value of  $\tau_{\rm {\scriptscriptstyle GL}}^{-1}=\alpha (T-T_{\rm c})$
choosing $K^{(L)}(\epsilon)=\alpha/2\epsilon$. We consider the system invariant under the time reversal symmetry. So we should choose $K^{(T)}=0$ because otherwise this term would give the unnatural contribution to $\tau^{-1}_{\rm {\scriptscriptstyle GL}}$ changing its sign when, e.g., we reverse the direction of all currents in the system.

The important question is how this formalism describes the phase transition interface in the density-matrix space, see Fig.\ref{fig3}. The parameter $\tau_{\rm {\scriptscriptstyle GL}}^{-1}$ should not depend upon the choice of $F_{\rm L}^{(0)}$, thus $F_{\rm L}^{(0)}$ and $F_{\rm L}^{(0')}$ belonging to the interface should satisfy the relation
\begin{gather}\label{eq:condition}
    \int_{-\omega_D}^{\omega_D} \frac{d\epsilon}{2\epsilon}  \left[\tilde F_{\rm
L}^{(0')}(\epsilon)-\tilde F_{\rm L}^{(0)}(\epsilon)\right]=0,
\end{gather}
where $\omega_{\rm {\scriptscriptstyle D}}$ is the Debye energy.

The microscopic derivation of the GL action shows that Eq.~\eqref{eq:condition} can
be interpreted as the integral representation of the electron-phonon interaction
constant, $\lambda$, giving the BCS-superconductivity:
\begin{gather}
\int_{-\omega_D}^{\omega_D} \frac{d\epsilon}{2\epsilon} \tilde F_{\rm
L}^{(0)}(\epsilon)=\int_{-\omega_{\rm {\scriptscriptstyle D}}}^{\omega_D}
\frac{d\epsilon}{2\epsilon} \tanh(\epsilon/2T_{\rm c})=\frac1{\nu\lambda},
\end{gather}
where $T_{\rm c}=2\gamma\omega_{\rm {\scriptscriptstyle D}}/\pi e^{-1/\nu\lambda}$
and $\gamma=e^C$, with $C=0.577\ldots$ being the Euler constant.

In the Fourier space $L_{\rm {R(A)}}^{-1}=\pm i\omega+\tau_{\rm {\scriptscriptstyle
GL}}^{-1}+Dq^2$.
The Keldysh component $L_{\rm K}^{(-1)}$ in an equilibrium should satisfy the
relation following from the fluctuation-dissipation theorem
(FDT)~\cite{Levchenko-Kamenev}:
\begin{gather}
{[L^{(-1)}]_{\rm K}=B_{\rm {\omega}} (L_{\rm R}^{-1}-L_{\rm A}^{-1})_\omega,}
\end{gather}
where $B_\omega=\coth \omega/2T$ is the equilibrium
distribution function of the complex $\Delta$-field. The similar relation holds for
the out of the equilibrium state where the gradients of $B$ with respect the ``the
center of mass'' Wigner transformation variables are irrelevant, which is the case
we consider. The main (infrared) frequency scale of the Landau theory is $\tau_{\rm
{\scriptscriptstyle GL}}^{-1}$ and $Dq^2\sim \omega\sim \tau_{\rm
{\scriptscriptstyle GL}}^{-1}$. In an equilibrium $\tau_{\rm {\scriptscriptstyle
GL}}^{-1}\sim (T-T_{\rm c})\ll T_{\rm c}$  and, therefore, $2i\omega B_\omega\to 4i
T_{\rm c}$. In the out of the equilibrium state we should choose $\tau_{\rm
{\scriptscriptstyle GL}}^{-1}$ smaller than any relevant energy scale of $B_\omega$.
Then we can also replace $2\omega B_\omega$ by
\begin{gather}
{\lim_{\omega\to 0} 2\omega B_\omega\equiv T^*}
\end{gather}
and consider $T^*$ as the second effective temperature. So
\begin{gather}
{[L^{(-1)}]_{\rm K}\equiv4iT^*.}
\end{gather}

A recipe for constructing the stationary nonequilibrium distribution function is sketched in Fig.\ref{fig3}b.  The superconductor is connected to the electron reservoirs through the wires with the normal resistances $R$ and $R'$ [e.g., the disordered quasi-1D normal metal wires]. The reservoir biases and temperatures are, in general, different. If the wire resistance satisfies the relation, $R\gg 1/L\sigma_{\rm {\scriptscriptstyle N}}$ [$L$ and $\sigma_{\rm {\scriptscriptstyle N}}$ are the S-diameter and the normal conductivity of the S-material] then the current coming from the wires spreads over the island that ensures the weakness of $\tilde F_{\rm {e(h)}}$ ($\varphi$) gradients. The supercurrent $j_s$ is much smaller than the critical current if $\xi T_{\rm {eff}}/(L\langle\Delta_1\rangle )\gg L\sigma_{\rm {\scriptscriptstyle N}} R$
[the normal current -- supercurrent conversion at the NS interface is considered in Ref.\cite{PRBvariant} using the generalized boundary conditions \cite{Zaitsev_RO}, which take into account the proximity effect].
We neglect the Coulomb blockade effects considering the transmission probability (per channel) between the island and the wires to be close to unity. Normal wire -- superconductor island structures used in the experiments \cite{Baturina} satisfy most of these conditions. Solving the normal state KE, $D\partial_{\mathbf{r}}^2F_{\rm {e(h)}}=0$, we find that on the island, $F_{\rm {e(h)}} =\sum_{\rm n}
p_{\rm n}\tanh[(E-V_{\rm n})/2T_n]$, $\sum_{\rm n} p_{\rm n}=1$, where $V_{\rm n}$
is the voltage at the terminal $n$ and $p_{\rm n}$ is determined by the resistivity
of the wires [the exemplary system size is smaller than the quasiparticle inelastic
length scale].  For the two terminal case, $p_2=R/R_\Sigma$, where
$R_{\Sigma}=R+R'$. Similar structure of $p_n$ holds for a multiterminal case.

To illustrate the developed approach, we consider an exemplary system  with
$p_2=p_1=1/2$.  Then
\begin{gather}\label{eq:F_fig}
\tilde F_{\rm {e(h)}}=\frac12\left[\tanh\frac{E\pm V/2}{2T}+\tanh\frac{E\mp V/2
}{2T}\right],
\end{gather}
where $\delta\varphi=0$ \cite{footnote:imbalance}, $T^*=T_{\rm eff}$, and
\begin{gather}
{T_{\rm eff}=T\cosh^2(V/4T)}\,  ,
\\\label{eq:b_island}
{\tau_{\rm {\scriptscriptstyle GL}}^{-1}=\alpha\,T_{\rm eff}\left\{\Re \mathrm{H}_{-\frac12+iu}+2\ln2+\ln( T/{T_{\rm c}})\right\},}
\\
{b=-\frac{T_{\rm eff}}{2\pi^3 T} \Re\Psi^{(2)}\left(1/2+iu\right).}
\end{gather}
Here $u=V/{4\pi T}$, $\mathrm{H}$ is the Harmonic number and $\Psi$ is the  Digamma
function.
Importantly, $b$ is very sensitive to the degree of nonequilibrium, see
Fig.\ref{fig3}: it changes sign at $u\approx 0.3$ remaining negative at larger $u$
that signals of the possible instability [in agreement with \cite{Nazarov}] and
requires keeping $\sim\Delta^6$-terms in the action \cite{PRBvariant}. While
$u\lesssim 1$, $\tau_{\rm {\scriptscriptstyle GL}}^{-1}\approx [\theta
u^2+\ln(T/T_{\rm c})]\alpha\,T_{\rm eff}$ with $\theta=7 \zeta[3] \approx 8.4$. In
Fig.\ref{fig3}, $\langle\Delta\rangle\sim\sqrt{V_{\rm c}-V}$, with $V_{\rm c}\approx
4\pi T\sqrt{\ln( {T_{\rm c}}/T)/\theta}$.  Taking, e.g., $T=1.3 T_{\rm c}$ in the
reservoirs and  $V=2 T_{\rm c}$ we get: $T_{\rm eff}/T\sim 0.87$,
$\tau_{\rm{GL}}/\tau_{\rm {\scriptscriptstyle GL}}^{\rm (eq)}\sim 0.69$ and then the
fluctuation susceptibilities, see Eq.\eqref{eq:xi}: ${\chi}/{\chi^{(\mathrm{
eq})}}\sim 0.6$.

{To conclude, we have constructed the nonequilibrium GL theory on the symmetry
grounds under the condition that the kinetics of the high-symmetric phase is
established.
The coefficients of the nonequilibrium GL functional, which are the constants in an
equilibrium, become strongly dependent on the external drive in an nonequilibrium
state.   In particular, the coefficient at the forth order (in the order parameter)
term can change its sign at large driving forces; this would signal the onset of the
instability which requires the higher order expansion. The energy parameter $T_{\rm
eff}$ replacing the equilibrium temperature $T$, is now a nonlinear function of the
bias voltage.}
We have demonstrated that the fluctuation corrections to observable quantities, e.g., to
the magnetic susceptibility, in a superconducting island get strongly renormalized
and become the singular functions of $\sqrt{V-V_{\rm c}}$ when out of the
equilibrium rather than being functions of $\sqrt{T-T_{\rm c}}$ in an equilibrium.
Accordingly, the order parameter vanishes like $\sqrt{V_{\rm c}-V}$ in the out of
the equilibrium state replacing its $\sqrt{T_{\rm c}-T}$-dependence of an
equilibrium state.

We thank T.\,Baturina, Yu.\,Galperin, N.\,Kopnin and  R.\,Fazio for helpful
discussions. The work was funded by  RFBR,   the Deutsche Forschungsgemeinschaft GK
638, and by the U.S. Department of Energy Office of Science through the contract
DE-AC02-06CH11357.


\begin{thebibliography}{99}
\bibitem{footnote:imbalance} The quasiparticle imbalance effect~\cite{Kopnin_book},
causes a change in the electric potential, $\varphi=\varphi_0+\delta\varphi $,
$\varphi_0=\varphi(\Delta=0)$.  Generally, $\delta\varphi$, depends on
$\langle\Delta_1\rangle$ and $\tilde F_{T}$ [$\delta\varphi\sim \int_\epsilon
N_\epsilon \tilde F_T(\epsilon)$, where $N$ is BCS DoS]. The same applies to  $j_s$.
As we restrict ourselves to superconducting sytems with small gradients of $\varphi$
($j_s$) and strong nonequilibrium ($|V|\gg\langle\Delta_1\rangle$), Fig.\ref{fig3}, so $\delta\varphi\sim o(\langle\Delta_1\rangle/T_{\rm eff})$ and
the NGL-coefficients should be calculated using $\tilde F_{e(h)}|_{\delta\varphi=0}$
[$\delta\varphi$ is important, e.g., for the proximity coupled
S-islands \cite{PRBvariant}.]
\end{thebibliography}
\end{document}